\documentclass[aps,prl,reprint,superscriptaddress]{revtex4-2}
\usepackage{graphicx}
\usepackage{amsmath}
\usepackage{bm}
\usepackage{hyperref}

\begin{document}

\title{Wettability from Diffusion: A Universal Molecular Scaling Law}

\author{Lorenzo Agosta}
\affiliation{Department of Chemistry, Princeton University, Princeton, NJ 08544, USA}
\affiliation{Department of Chemistry, \AA{}ngstr\"om Laboratory, Uppsala University, 75121 Uppsala, Sweden}

\begin{abstract}
Quantifying wettability at the nanoscale remains challenging, as macroscopic contact-angle measurements fail to capture the molecular interactions that define hydrophilic and hydrophobic behavior. We derive an analytical relation linking the water contact angle to the lateral diffusion of interfacial molecules, establishing a quantitative connection between microscopic dynamics and macroscopic wettability. Molecular dynamics simulations confirm that the ratio of interfacial to bulk diffusion uniquely determines the contact angle across the full hydrophilic-hydrophobic spectrum. This diffusion-based formulation eliminates the need for droplet geometries or free-energy sampling, enabling quantitative assessment of wetting directly from short \emph{ab initio} molecular dynamics trajectories. The approach provides a universal and efficient route to evaluate surface affinity in reactive, defective, or confined environments.
\end{abstract}

\maketitle

\section{Introduction}

Wettability governs how liquids interact with solid surfaces and plays a central role in phenomena ranging from catalysis and corrosion to energy conversion, adhesion, and biomolecular recognition. It is traditionally quantified by the macroscopic contact angle (CA), defined by Young’s equation \cite{young}, which balances the surface free energies of the solid–liquid, solid–vapor, and liquid–vapor interfaces. Despite its simplicity, this macroscopic descriptor suffers from several source of discrepancies such as hysteresis~\cite{Machata2022,Shi2018,Andreotti2020}, dependency on surface defects and impurities which affect the droplet baseline used to extract the angle~\cite{Kumar2007,Kaplan2013,Bonn2009}. The  applicability of the CA becomes even more challenging at the nanoscale \cite{wet_nano} where the wetting is controlled by intrinsic molecular interactions, and  in reactive or confined systems where droplet formation is impossible or ill-defined, the contact angle loses physical meaning\cite{Jahnert2008}.

Alternative microscopic measures of wettability have been proposed to overcome these limitations. Hydrodynamic approaches based on slip length or interfacial friction \cite{Huang2008,Sendner2009,Oga2019} provide useful correlations but remain indirect, while spectroscopic and statistical methods such as interfacial hydrogen-bond analysis \cite{Pezzotti2021} or density-fluctuation metrics \cite{Patel2010,Patel2011,Rego2022,Jamadagni2011} offer qualitative insight but lack quantitative transferability. From the computational perspective, molecular dynamics (MD) simulations allow explicit access to interfacial structure and energetics, yet droplet-based CA calculations are computationally demanding, requiring large system sizes and nanosecond to microsecond sampling \cite{Jiang2019,Limmer2013}. These challenges are even more severe for ab initio molecular dynamics (AIMD), where electronic effects describing water reactivity and polarization upon adsorption must be included explicitly \cite{Banuelos2023}.

At the molecular level, several studies have revealed that water dynamics at solid interfaces encode information about the local hydrophobicity or hydrophilicity. Hydrophilic surfaces constrain molecular mobility through strong adsorption and hydrogen bonding\cite{drossel,graphene_ox,water_silica_diff,diffusion_interface_rev,Agosta2017}, whereas hydrophobic walls enhance translational diffusion by reducing interfacial friction \cite{rosky94,gonzales08,cicero,agosta25,agosta_ceo2water}. These observations suggest that the dynamic response of interfacial water could serve as a direct probe of wettability.


Building on this premise, we introduce a diffusion-based framework that quantitatively links interfacial molecular motion to macroscopic wetting. Specifically, we derive an analytical relation connecting the lateral diffusion of water molecules adjacent to a surface to the macroscopic contact angle. This scaling law provides a quantitative connection between molecular mobility and adhesion free energy, validated through molecular dynamics simulations across the full hydrophilic–hydrophobic spectrum. The formulation eliminates the need for droplet geometries or free-energy integration, allowing wettability to be inferred directly from short equilibrium simulations, including those performed within ab initio frameworks. This approach establishes diffusion as a universal, physically transparent descriptor of wetting, bridging microscopic dynamics with macroscopic surface thermodynamics.

\section{Results}
\textbf{Contact angles and water structure} To establish a quantitative link between diffusion and wettability, we first examined the structure of liquid water near model surfaces with tunable hydrophilicity. Classical molecular dynamics (MD) simulations were carried out using the flexible SPC/Fw water model \cite{spcf_model} in contact with smooth Lennard–Jones (LJ) walls. The LJ interaction strength, $\varepsilon$, was varied between 0.5 and 1.5 kcal/mol, spanning the full range from strongly hydrophobic to fully hydrophilic regimes. All systems were equilibrated at 300 K, and periodic boundary conditions were applied parallel to the wall.

\begin{figure*}[ht]
\centering
\includegraphics[width=0.99\textwidth]{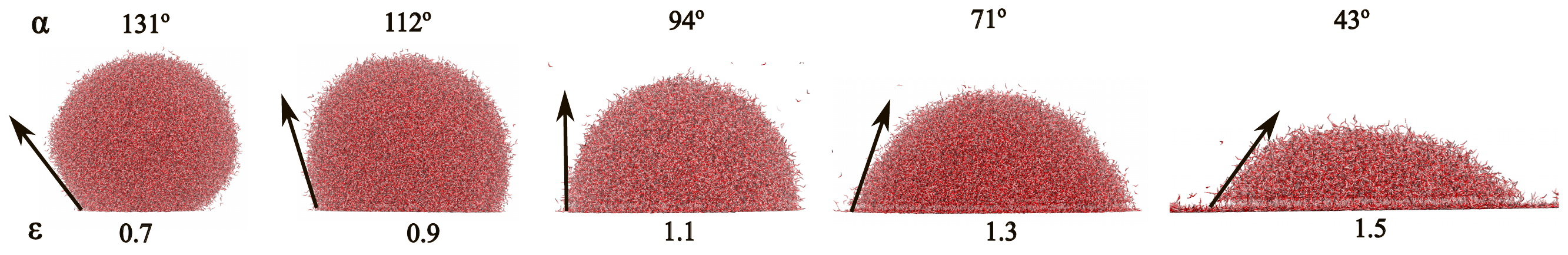}
\caption{Water droplets on the increasing hydrophilic LJ surfaces. Each droplet represents the overlap of 100 snapshots. The water contact angle is estimated with the methodology described in the method section.}
\label{fig:wca}
\end{figure*}

The resulting equilibrium morphologies display the expected monotonic evolution of wettability with $\varepsilon$. At low $\varepsilon$, the water droplet exhibits large contact angles, while increasing wall–water attraction gradually enhances spreading, yielding smaller contact angles, see Figure \ref{fig:wca}. For each system, the contact angle $\alpha$ was extracted from the density profile of water molecules as described in the Method section.

\begin{figure*}[ht]
\centering
\includegraphics[width=0.99\textwidth]{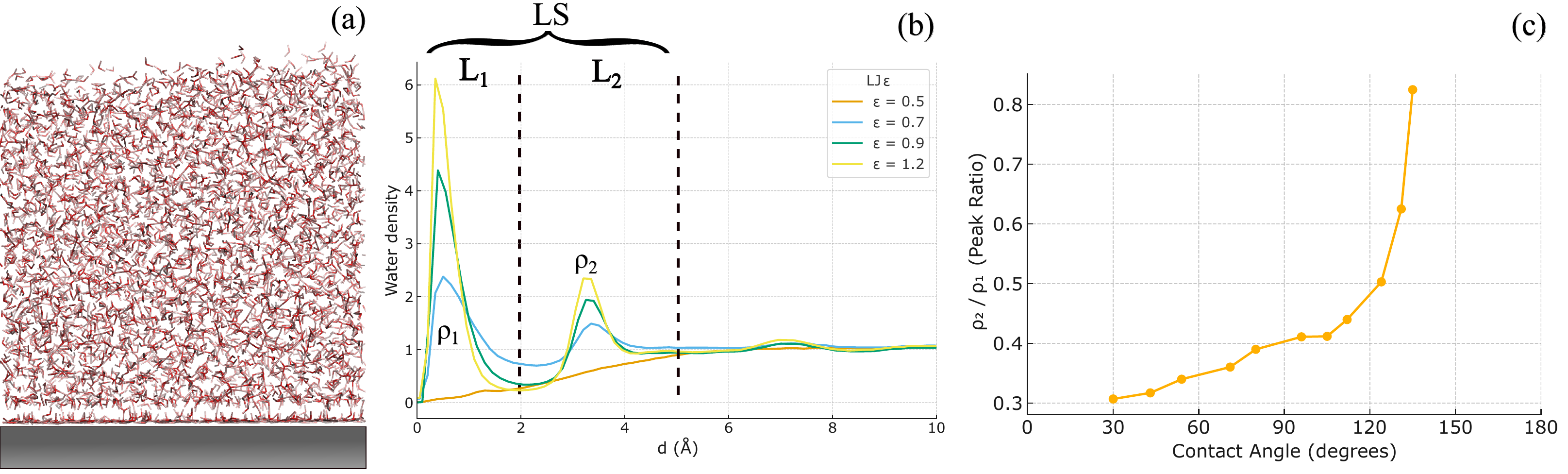}
\caption{ (a) Representation of the water film on top of the adsorbing LJ wall ($\epsilon =1.5$). (b) Water oxygen density profiles for different LJ $\epsilon$ parameters. (c) Water density peak ratio as a function of the relative contact angles. }
\label{fig:densities}
\end{figure*}

The microscopic water structure normal to the interface was analyzed for the simulated water films (Fig. \ref{fig:densities}a). The oxygen density profile, $\rho(z)$, averaged over the simulation time is displayed in Figure \ref{fig:densities}b. Hydrophilic surfaces show pronounced layering, characterized by two main density maxima located at approximately 1 and 3~Å from the wall, in agreement with previous observations of structured interfacial water \cite{chandler_wall_ideal,Sendner2009}. For $\varepsilon < 0.6$, the layering disappears, indicating a purely hydrophobic regime where adsorption is negligible and water retains bulk-like structure. The effect of the wall extends over roughly 5~Å, which defines the effective thickness of the liquid–solid interfacial region, $LS{_i}$. A direct correlation emerges between the ratio of the first and second density peaks ($\rho_1/\rho_2$) and the macroscopic contact angle: $\rho_1/\rho_2 \approx 0.4$ corresponds to $\alpha \approx 90^\circ$, identifying the hydrophilic–hydrophobic crossover point.

\textbf{Derivation of the diffusion–wettability relation} To establish a quantitative law connecting diffusion and wettability, we start from the Young–Dupré equation \cite{Schrader1995}

\begin{equation} \label{dufre}
    cos\alpha = \frac{-W_{LS}}{F_{LV}} -1
\end{equation}

where $W_{LS}$ is the work of adhesion between the liquid and solid, and $F_{LV}$ is the surface free energy of the liquid–vapor interface.


Several approaches have been proposed to evaluate the adhesion free energy 
\( W_{LS} \) from molecular dynamics simulations. 
The most rigorous route is thermodynamic integration, in which the reversible work 
required to separate the solid–liquid and liquid–vapor interfaces is sampled along 
a controlled reaction coordinate~\cite{Jiang2019}. 
Although formally exact, this method demands extremely long simulations and extensive 
sampling to achieve statistical convergence. 
A second approach approximates \( W_{LS} \) from the fluctuations of the interaction 
energy between liquid molecules and the solid surface~\cite{Saito2021}. 
This method also suffers from large statistical noise, as interfacial energy 
fluctuations are inherently broad and slow to converge. 
Finally, a more empirical approximation estimates \( W_{LS} \) from the enthalpic 
contribution of an ordered, ice-like water layer adsorbed on the solid surface~\cite{lopez16}. 
While computationally simple, this approach neglects entropic effects and cannot 
capture the dynamic nature of liquid interfaces. 

All these methods share the common goal of quantifying the free energy associated 
with liquid molecules adsorbed near the solid interface. 
In line with this philosophy, the approach introduced in this work seeks to evaluate 
the same quantity through an alternative, dynamic route: by linking \( W_{LS} \) to 
the molecular diffusion of water at the interface. 
We interpret \( W_{LS} \) as the free energy required to form the structured water 
layer adjacent to an adsorbing wall ($L_1$), starting from the reference case of a 
non-interacting surface. 
Let \( F_{LS_i} \) denote the total free energy associated with forming the interfacial 
water structure within the first 5~\AA{} of the wall, relative to the non-interacting condition. 
To isolate the contribution of the first adsorbed layer only, the free energy of the 
second liquid layer (\( F_{L_2} \)) and that of water in contact with a purely repulsive 
(non-interacting) wall (\( F_{L_{nni}} \)) must be subtracted. 
The resulting expression defines the effective adhesion free energy as:

\begin{equation} \label{F1}
   F_{L_1} = F_{LS_i} - F_{L_2} - F_{L_{nni}} = W_{LS}.
\end{equation}



We assume these free energy differences are proportional to the local activation free energy for molecular diffusion. 
As mentioned in the introduction this heuristic assumption is plausible as it was observed that hydrophilicity is proportional to reduced water diffusion (and vice-versa for the hydrophobic case). In addition, the activation energy was proven to be equal to the local free energy for hard-sphere liquids \cite{dzugutov_HF}, and this relation was extend to simple liquids \cite{dzugutov1,Agosta2024} and water confined by hydrophobic walls \cite{agosta25}.
The Arrhenius relation for diffusion states:

\begin{equation} \label{Arrenius}
  F= -kT ln \frac{D}{D_0} 
\end{equation}

where $k$ is the Boltzmann factor, $T$ is the temperature of the system and where $D$ and $D_0$ denote the diffusion coefficients of a local state and its reference, respectively. 

Using Eq. \ref{Arrenius} to describe the terms in Eq. \ref{F1} one obtains

\begin{equation}\label{F2}
  F_{L_{2}}= -kT ln{ \frac{D_{L_{2}}} {D_{b}}   } , \, \, \, F_{LS_i}= -kT ln{\frac{D_{LS_i}}{D_{L_{nni}}}  }
\end{equation}

\begin{equation} \label{Fnni}
  F_{L_{nni}}= -kT ln {\frac{D_{b}}{D_{L_{nni}}}  } 
\end{equation}

Eq. \ref{Fnni} was proven to be exact for water in contact with a repulsive wall \cite{agosta25}.
Combining together Eq. \ref{F1},\ref{Arrenius}, \ref{F2} and \ref{Fnni} the expression for the adhesion energy becomes:

\begin{equation}
  W_{LS}= -kT ln {\frac{D_{LS_i}}{D_{L_{2}}}  } 
\end{equation}

and finally Eq. \ref{dufre} can be rewritten as:

\begin{equation}  \label{diff-ca}
cos\alpha = \frac{ ln {\frac{D_{LS}}{D_{L_{2}}}} }{ ln{\frac{D_{b}} {D_{LV} } }  }   -1
\end{equation}

Since $D_b$ and $D_{LV}$ are intrinsic properties of the water model at a given temperature, Eq. \ref{diff-ca} provides a parameter-free relation between measurable molecular diffusion and macroscopic wettability. Moreover, $D_{LV}$ can be estimated from $D_{L_{nni}}$ as these quantities are equal,  as shown in Fig. S1 in SI.

\begin{figure}[ht]
\centering
\includegraphics[width=0.95\columnwidth]{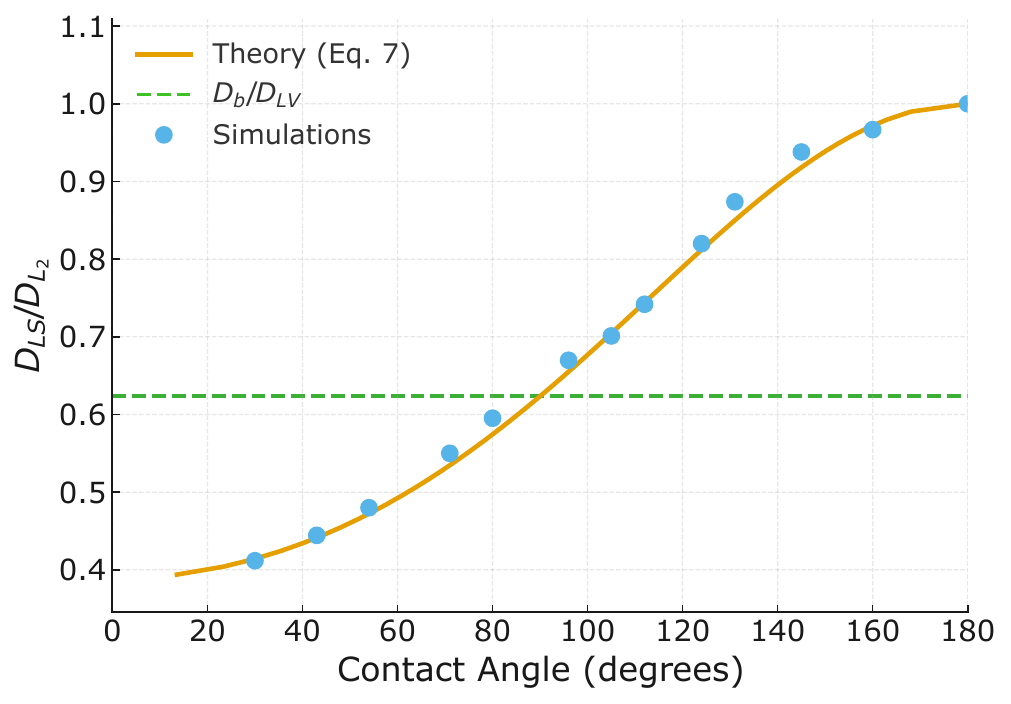}
\caption{Ratio of lateral diffusion coefficients at the liquid-solid interface  ($D_{LS_i}$) and the second water layer ($D_{L_2}$)  as a function of contact angle. Simulation results for Lennard-Jones surfaces are compared with the theoretical prediction from Eq.~7, showing excellent agreement across the full wettability range.}
\label{fig:diffusion}
\end{figure}

\textbf{Validation and experimental accessibility} The predictions of Eq. \ref{diff-ca} were tested against MD simulations covering the full hydrophilic–hydrophobic spectrum. Figure~\ref{fig:diffusion} reports the ratio $D_{LS}/D_{L2}$ as a function of the computed contact angle. The analytical law reproduces the simulation results with quantitative accuracy across all regimes. In the hydrophobic limit, where $D_{LS}=D_{L2}$, the model yields $\alpha=180^{\circ}$, corresponding to a non-wetting surface. This implies that the $L_1$ and $L_2$ are not distinguishable (Figure \ref{fig:densities}b for $\epsilon = 0.5$) in the case of super hydrophobicity. To note that for $ 0.5<\epsilon<0.65$, $D_{LS}$ becomes larger than $D_{L2}$ and in  Eq. \ref{dufre} $W_{LS}$ must be taken with a change in sign.
Conversely, for $D_{LS}=D_b$, the predicted $\alpha=90^{\circ}$ represents a neutral interface. Intermediate cases capture the continuous transition between these limits, including the superhydrophilic regime ($\alpha < 30^{\circ}$). To note that the value $\alpha =90^0$ is reached for $D_{L_{LS_i}}/D_{b}$ = 0.6, meaning that the the surface displays hydrophobicity even if the interfacial diffusion is reduced with respect to the bulk (up to $D_{L_{LS_i}}/D_{b}$ = 1, see Fig. S3 in SI).

Although $D_{LS}/D_{L2}$ is straightforward to compute in MD simulations, experimental determination of such nanoscopic diffusion profiles might be challenging. However, recent advances in nuclear magnetic resonance (NMR) spectroscopy now enable direct probing of interfacial water dynamics with \AA{}ngstr\"om-level sensitivity. In particular, Overhauser dynamic nuclear polarization (ODNP) spectroscopy using stable radical markers immobilized on surfaces has been shown to resolve interfacial water diffusivity and its modulation by surface chemistry~\cite{Overhauser,Overhauser2}. Moreover, relaxation-time analysis of confined water systems has revealed exchange between surface-bound and bulk-like water states~\cite{gravelle25}, providing an experimental route to extract the $D_{LS}/D_{L2}$ ratio. These developments make it feasible to test Eq. \ref{diff-ca} experimentally and to define wettability directly from interfacial water mobility.

\textbf{Relation to Young’s equation and physical interpretation} Finally, the proposed diffusion-based law can be connected back to the classical Young equation,

\begin{equation}
    cos\alpha = \frac{F_{SV}-F_{LS}}{F_{LV}} 
\end{equation}

where $F_{SV}$ and $F_{SL}$ denote the solid–vapor and solid–liquid surface free energies, respectively.
If we assume that the liquid–vapor interface behaves as a non-interacting limit, then $F_{LV} = F_{L_{nni}}$ (see Fig. S1 in SI), from Eq. \ref{dufre} and \ref{F1} it follows that the free energy of the solid–vapor interface can be identified with that of the second water layer, $F_{SV} \approx F_{L_2}$. Intriguingly, $F_{L_2}$ is considered the equilibrium state between the an adsorbing solid surface and the liquid phase in the Brunauer, Emmett, and Teller (BET) theory \cite{BET}. This is again in agreement with the observation that  ${L_2}$ represents the solid-vapor equilibrium, and that $-kT ln{ \frac{D_{L_{2}}} {D_{b}}}$ is a good estimator of the solid-vapor surface free energy.   This interpretation reveals that the microscopic diffusion ratios encode the same free-energy differences appearing in the macroscopic Young formulation.
In this picture, interfacial diffusion plays the role of an effective thermodynamic probe: reduced mobility corresponds to higher adhesion energy and smaller $\alpha$, while enhanced diffusion reflects weaker binding and larger $\alpha$. Thus, the molecular diffusion of interfacial water can be viewed as the dynamic counterpart of Young’s surface free-energy balance.

\section{Discussion and Conclusions}
We demonstrate a direct, quantitative relation between interfacial diffusion and surface wettability. The proposed scaling law bridges microscopic dynamics and macroscopic wetting without recourse to droplet geometries or thermodynamic integration. The method requires only short equilibrium simulations and is therefore directly applicable within \emph{ab initio} molecular dynamics, where long sampling times are prohibitive.

The physical origin of this relation lies in the interplay between interfacial entropy and adhesion free energy. Hydrophilic surfaces impose structural ordering and restrict translational mobility, reducing entropy and diffusion, while hydrophobic surfaces minimize adsorption and enhance molecular freedom \cite{Agosta2024}. The derived scaling law therefore captures, in compact analytical form, the thermodynamic balance that defines wettability. 

This approach establishes diffusion as a universal, dynamics-based descriptor of wettability. It enables predictive assessment of hydrophilic and hydrophobic behavior in systems where conventional contact-angle methods fail, including reactive oxides, coated perovskites, and nanoconfined fluids.

\section{acknowledgments}
The author acknowledges support from the Swedish Research Council (VR International Postdoc Grant) and Princeton University. Computational resources were provided by the National Supercomputer Centre of Sweden (SNIC).

\section{Methods} 

\subsection*{Simulations details}

The molecular dynamics simulations were performed with LAMMPS software\cite{lammps} at 300K in a NVT ensemble using a SPC flexible model for water-water interactions \cite{spcf_model} and with an integration time step of 1.0 fm. Two different periodic boxes were considered. The simulations related to the water droplet formation were set by considering a periodic box (in xy) of $140x140x140$ \AA $^3$. A Lennard-Jones wall ($\sigma=1.0$ \AA, $cutoff=10.0$ \AA) was placed perpendicular to the z direction at the -1 \AA \, from the $z=0$ \AA of the box. The LJ $\epsilon$ parameter of the wall was varied starting from 0.5 kcal/mol (non interacting condition) to 1.5 kcal/mol (fully hydrophilic condition) with a 0.1 increment. 2133 water molecule in a cubic box of $40x40x40$ \AA $^3$ (relative to liquid water density at 300K and 1 atm) were placed at 2 \AA from the wall along the z direction. In order to avoid the water molecules transitioning to the vapor phase to interact with the z image of the simulation a second repulsive wall was set at 80 \AA from the the first wall along the z direction. The simulations were run for 20 ns (for each LJ interactive parameter) .
The simulations related to the water film were set using a periodic box (in xy) of $50x50x300$ \AA $^3$. The same Lennard-Jones wall used for the droplet simulation was adopted. A film of 4167 water molecules was place in a  $50x50x50$ \AA $^3$ cubic box at  2 \AA \, distance from the wall.
The simulations were run for 2 ns.

\subsection*{Lateral Mean Square Displacement}

The components of the diffusion coefficient parallel to the wall are evaluated from the mean-square displacement (MSD) corrected by the probability for a particle to remain within a layer during the time interval $t$ \cite{Liu2004, Agosta2017,supercooled}.

\begin{equation} \label{msd} 
  D_{xy} (d) = \frac{1}{4} \lim_{t \to \infty}
  \frac{\text{MSD}_{xy} (d, t)}{t P(d, t)} 
\end{equation}

where the MSD is defined in a layer at the distance $d$ from the wall. $P(d, t)$ is the probability that a particle stays in the layer within time interval $t$. In all layers the latter time was found to be sufficient for the particles to exhibit long enough linear regime of MSD needed for the reliable calculation of the diffusion coefficient.

\subsection*{Contact Angle determination}

In order to evaluate the contact angle  the following formula was used:

\begin{equation} \label{CA} 
  cos\alpha = \frac{d^2-c^2}{d^2+c^2} 
\end{equation}

where $d$ is the height of the droplet from the wall and $c$ is its radius at the baseline. In order to estimated $d$ and $c$ the averaged density profile of water molecules along the z direction and water molecules lying in the xy place in contact with wall were calculated.

\bibliographystyle{apsrev4-2}
\bibliography{wetting_refs}

\end{document}